\documentclass{ifacconf}
\usepackage{graphicx}      % include this line if your document contains figures
\usepackage{natbib}        % required for bibliography
\usepackage{amsmath} % assumes amsmath package installed
\usepackage{amssymb}  % assumes amsmath package installed
\usepackage{amsfonts}
\usepackage{nicematrix}
\usepackage{color}
\usepackage{array}
\usepackage{booktabs}
\usepackage{url}
\usepackage{caption}
\usepackage{todonotes}

\newtheorem{problem}{Problem}

\begin{document}
\begin{frontmatter}

\title{Safety-Critical Control for Discrete-time Stochastic Systems with Flexible Safe Bounds using Affine and Quadratic Control Barrier Functions}%\thanksref{footnoteinfo}} 
% Title, preferably not more than 10 words.

% \thanks[footnoteinfo]{Sponsor and financial support acknowledgment goes here. Paper titles should be written in uppercase and lowercase letters, not all uppercase.}

\author[First]{Sotaro Fushimi} 
\author[Second]{Kenta Hoshino} 
\author[Third]{Y\^uki Nishimura} 

\address[First]{Faculty of Engineering, Kyoto University, Yoshida-honmachi, Sakyo-ku, Kyoto 606-8531, Japan\\
   (e-mail: s-fushimi@sys.i.kyoto-u.ac.jp).}
\address[Second]{Department of Informatics, Kyoto University, Yoshida-honmachi, Sakyo-ku, Kyoto 606-8531, Japan \\
   (e-mail: hoshino@i.kyoto-u.ac.jp)}
\address[Third]{Graduate School of Science and Engineering, Kagoshima University, 1-21-40, Korimoto, Kagoshima 890-0065, Japan \\
   (e-mail: yunishi@mech.kagoshima-u.ac.jp)}

 \begin{abstract}                %Abstract of 50--100 words
This paper presents a safe controller synthesis of discrete-time stochastic systems using Control Barrier Functions (CBFs).
The proposed condition allows the design of a safe controller synthesis that ensures system safety while avoiding the conservative bounds of safe probabilities.
In particular, this study focuses on the design of CBFs that provide flexibility in the choice of functions to obtain tighter bounds on the safe probabilities.
Numerical examples demonstrate the effectiveness of the approach.
\end{abstract}

\begin{keyword}
Robustness, Lyapunov Methods, Disturbance Attenuation
\end{keyword}

\end{frontmatter}
%===============================================================================
\section{Introduction}
Safety is a crucial aspect of automation deployment and is typically characterized by the forward invariance of a specified safe set. 
To achieve forward invariance, Lyapunov-like methods using barrier functions and, more recently, Control Barrier Functions (CBFs) have been developed (see~\cite{Ames:19} for the review of CBFs).
CBFs provide criteria for the design to ensure the forward invariance in a prescribed safe set.
While these approaches are highly effective for deterministic systems—those free from uncertainties—they encounter significant challenges when applied to stochastic systems. 
In particular, when disturbances with infinite tails, such as Gaussian noise, are considered, synthesis methods for deterministic systems become inapplicable.
The unbounded nature of such disturbances makes it difficult to guarantee the forward invariance of safe sets using barrier approaches over an infinite time horizon as shown in \cite{so2023almost}.

Given the difficulty in ensuring the forward invariance with probability one,
one of the promising approaches to the safety of stochastic systems is to focus on the exit probability over a finite time horizon.
% As forward invariance with probability 1 is generally unattainable, researchers have focused on verifying the exit probability over a finite time horizon. 
One common approach involves finding nonnegative supermartingales of the system state, which serve as analogs of Lyapunov functions (\cite{kushner1966finite,kushner1967stochastic,prajna2007framework,steinhardt2012finite,santoyo2021barrier}). 
\cite{kushner1966finite, kushner1967stochastic} provide upper bounds on the probability that the values of nonnegative functions, serving as Lyapunov or barrier functions, ever exceed a specified threshold by utilizing the supermartingale property.
Building on this fundamental result, various methods for constructing supermartingales have been proposed.
\cite{steinhardt2012finite} developed a semidefinite programming approach for polynomial systems with safe sets defined by quadratic functions.
\cite{santoyo2021barrier} introduced a sum-of-squares formulation for constructing supermartingales for polynomial systems with polynomial barrier functions, enabling safe controller synthesis for affine-in-control systems that achieves a specified upper bound on exit probability.
\cite{cosner2023robust} adapted these barrier-based methods to CBF formulations for discrete-time systems, deriving risk probabilities when stochastic analogs of discrete-time CBFs (\cite{agrawal2017discrete}) are applied.

Previous works on discrete-time systems primarily focus on bounded safety regions with bounded barrier function values, typically for concave zeroing CBFs or convex barrier functions. 
Compared to affine constraints with respect to the control input used in continuous-time CBFs, discrete-time CBFs (\cite{agrawal2017discrete}) result in non-convex constraints with respect to the control input, except for concave CBFs. 
Combined with the fact that bounded barrier function values facilitate the construction of nonnegative supermartingales, bounded concave zeroing CBFs or bounded convex barrier functions have been the main focus in martingale-based stochastic safety studies. 
However, these assumptions limit applicability to realistic scenarios such as obstacle avoidance, where unbounded safe regions are often required (\cite{singletary2021comparative}). 
To the best of the authors' knowledge, the synthesis of safe controllers for unbounded safe regions remains largely unexplored for discrete-time systems, except in specific cases, such as when the time-step difference of CBF values is upper-bounded (\cite{cosner2024bounding}) or when barrier functions are learned (\cite{vzikelic2023learning}).

In this work, we focus on deriving conditions for the synthesis of safe controllers for discrete-time stochastic systems subject to Gaussian-distributed disturbances, considering both bounded and unbounded safe regions.
We first present a generalized condition for constructing nonnegative supermartingales and verifying the exit probability using Ville's inequality shown in~\cite{ville1939etude}. 
We then propose various sufficient conditions required to verify the exit probability for different classes of CBFs, extending the work of \cite{steinhardt2012finite}, \cite{santoyo2021barrier}, and \cite{cosner2023robust}.
Specifically, we propose conditions for polynomial bounded CBFs, and general affine and quadratic CBFs.
Finally, we formulate the safe controller synthesis as a (not necessarily convex) optimization problem by utilizing the safety verification method and provide a number of numerical examples to demonstrate the efficacy of our proposed method.
The contributions of this work include a comparison of different strategies for creating nonnegative supermartingales for bounded and unbounded safety regions, and the derivations of conditions for unbounded affine and quadratic CBFs, which, to the best of the authors' knowledge, have not been addressed previously.

The paper is organized as follows. 
Section~\ref{sec: preliminaries} provides preliminary definitions and results, including martingale properties and Ville's inequality.
Section~\ref{sec: prob} presents the problem formulation of this study.
Our main results are presented in Section \ref{sec: main}. 
Finally, numerical examples for affine and quadratic CBFs are provided in Section \ref{sec: num}.

\section{PRELIMINARIES}\label{sec: preliminaries}

The safe control problem addressed in this study is formulated for discrete-time stochastic systems, which is driven by Gaussian disturbances.
This section introduces notations used in this paper, followed by definitions and results of martingale properties of discrete-time stochastic processes that plays crucial roles to address the control problem.

Throughout this paper, the following notations are used.
The set of nonnegative integers is denoted by $\mathbb{Z}_{\ge 0}$.
The notations $\mathbb{R}$, $\mathbb{R}_{\ge 0}$, $\mathbb{R}^n$ represents the sets of real numbers, nonnegative real numbers, and the \(n\)-dimensional Euclidean space.
The notation $\mathbb{R}^{n \times m}$ denotes the set of matrices.
$0_m\in\mathbb{R}^{m}$ and $O_{m\times n}\in\mathbb{R}^{m\times n}$ denote the zero vector and the zero matrix, and $I_n \in \mathbb{R}^{n \times n}$ represents the identity matrix of dimension $n$.
The notation $\mathrm{diag}(a_1, \dots, a_n) \in \mathbb{R}^{n \times n}$ denotes the diagonal matrix with the $i$th diagonal entry $a_i$ ($i=1,\dots, n)$.
To set up stochastic settings, we use the notation $(\Omega, \mathcal{F}, \left\{ \mathcal{F}_k \right\}_{k \in \mathbb{Z}_{\ge 0}}, \mathbb{P})$ to denote a filtered probability space, where $\Omega$ is a sample space, $\mathcal{F}$ is a \(\sigma\)-algebra, $\{\mathcal{F}_k\}_{k \in \mathbb{Z}_{\ge 0}}$ is a filtration of $\mathcal{F}$, and $\mathbb{P}$ is a probability measure.
The Gaussian distribution on  $\mathbb{R}^n$ with the mean $\mu \in \mathbb{R}^n$ and the covariance matrix $\Sigma \in \mathbb{R}^{n \times n}$ is denoted by $\mathcal{N}(\mu, \Sigma)$.

We consider a scalar-valued stochastic process  $\left\{ W_k \right\}_{k \in \mathbb{Z}_{\ge 0}}$ on a filtered probability space $(\Omega, \mathcal{F}, \left\{ \mathcal{F}_k \right\}_{k \in \mathbb{Z}_{\ge 0}}, \mathbb{P})$.
The martingale properties are used to characterize CBFs in this study.
\begin{defn}
        A stochastic process $\{W_k\}_{k \in \mathbb{Z}_{\ge 0}}$ on a filtered probability space $\left( \Omega, \mathcal{F}, \left\{ \mathcal{F}_{k} \right\}_{k \in \mathbb{Z}_{\ge 0}}, \mathbb{P} \right)$ is a martingale if
        \begin{equation}
          \mathbb{E}[W_{k+1}|\mathcal{F}_k]=W_k \text{ a.s. }
        \end{equation}
        and is a supermartingale if
        \begin{equation}
          \mathbb{E}[W_{k+1}|\mathcal{F}_k]\leq W_k \text{ a.s.}
        \end{equation}
\end{defn}
The following result plays key roles in developing safe controller synthesis in this study.
This is due to~\cite{ville1939etude}.
\begin{lem}[Ville's inequality]
  If $W_k$ is a nonnegative supermartingale, then for all $\lambda>0$,
  \begin{equation}
  \label{eq:ville-inequality}
    \lambda \mathbb{P}\{{\rm sup}_{k\in\mathbb{Z}_{\ge 0}} W_k > \lambda\}\leq \mathbb{E}[W_0].
  \end{equation}
\end{lem}

\section{PROBLEM STATEMENT}\label{sec: prob}

This section formulates the safe control problem addressed in this study.
% Particularly, this study is interested in improving probability bounds of the safety provided in \cite{cosner2023robust}, which is introduced after presenting the formulation.
% \cite{cosner2023robust} の内容とTheoremが微妙に違うので，明らかにCosnerのimproveだと言うのは避けたいです．

Consider the following discrete-time control-affine systems on a filtered probability space $(\Omega, \mathcal{F}, \{\mathcal{F}_k\}_{k \in \mathbb{Z}_{\ge 0}}, \mathbb{P})$;
\begin{equation}
  x_{k+1}=f(x_k)+g(x_k)u_k+w_k, \label{eq: sys}
\end{equation}
where $x_k\in\mathbb{R}^n$ and $u_k\in\mathbb{R}^m$ are state and input at time step $k$, $f:\mathbb{R}^n\to\mathbb{R}^n$ and $g:\mathbb{R}^n\to\mathbb{R}^{n\times m}$ are continuous functions, 
and $w_k\in\mathbb{R}^n$ is a random disturbance whose probability distribution is given by the Gaussian distribution $\mathcal{N}(0, \Sigma)$ with $\Sigma$ being the covariance matrix.
For every $k$, $w_k$ is independent of $\mathcal{F}_k$ and is measurable with respect to  $\mathcal{F}_{k+1}$.
Furthermore, $u_k$ is assumed to be adapted to $\mathcal{F}_k$.
In the following, we use the notation $F(x_k,u_k)=f(x_k)+g(x_k)u_k$ to denote the drift term of system~(\ref{eq: sys}).

This study addresses the safety-critical control problem of stochastic discrete-time system~\eqref{eq: sys} where we will determine the controller $k_c:\mathbb{R}^n \to \mathbb{R}^m$ that yields the closed-loop system of~(\ref{eq: sys}),
\begin{equation}
\label{eq:closed-loop}
x_{k+1} = f(x_k) + g(x_k) k_c(x_k) + w_k.
\end{equation}
This study focuses on establishing conditions for $k_c$ to ensure that the trajectory $x_k$ of~(\ref{eq:closed-loop}) remains within a prescribed safe set.
We suppose that the safe set $\mathcal{C} \subset \mathbb{R}^n$ is given by a continuous function  $h:\mathbb{R}^n\rightarrow\mathbb{R}$ as follows:
\begin{equation}
  \mathcal{C}=\{x\in\mathbb{R}^n : h(x)\geq 0\}. \label{eq: safe set}
\end{equation}

The safety-critical control problem has been studied for deterministic systems, where the safety is typically characterized by the forward invariance; the state remains in the safe set $\mathcal{C}$ over the infinite horizon.
This study focuses on stochastic system~(\ref{eq: sys}), where such a forward invariance does not generally hold because of the Gaussian disturbance whose distribution has the unbounded support.
To formulate the safety-critical control problems in the stochastic setting,
we define the probability of the system exiting the safe set $\mathcal{C}$ specified by \eqref{eq: safe set} as follows:
\begin{defn}($K$-step safe in probability)
Given a controller $k_c$ and the initial state $x_0 \in \mathcal{C}$, system~(\ref{eq:closed-loop}) is \textit{\(K\)-step safe with probability $1-\epsilon$} for some $\epsilon \in (0,1)$ if
\begin{equation}
  P(x_0,K):=\mathbb{P} \left\{ x_{k} \notin \mathcal{C} \text{ for some } 0 \le k \le K \right\} \le \epsilon.\label{eq: Kstep exit}
% \mathbb{P} \left\{ x_{k} \notin \mathcal{C} \text{ for some } k \le K \right\} \le 1 - \epsilon.
\end{equation}
We refer to $P(x_0, K)$ as the \textit{\(K\)-step exit probability} of system~(\ref{eq:closed-loop}) from the safe set $\mathcal{C}$.
\end{defn}

The primary problem addressed in this study is formulated as follows:
\begin{problem}
\label{prob:problem-statementp}
Given system~(\ref{eq: sys}), design a feedback controller $k_c$ so that closed-loop system~(\ref{eq:closed-loop}) is \(K\)-step safe with probability $1-\epsilon$ for some $\epsilon \in (0,1)$.
\end{problem}

The safety threshold $\epsilon$ plays a crucial role in ensuring safety in Problem~\ref{prob:problem-statementp}, as it is preferable to minimize $\epsilon$.
% This study particularly focuses on developing a control design to achieve a tighter bound $\epsilon$ than those in the results of~\cite{cosner2023robust,cosner2024bounding}.
%より強いboundはメインのトピックではない（あくまでunboundedがメイン？）ので，もう少し弱い書き方をする．
A number of studies have been conducted to address Problem~\ref{prob:problem-statementp} (\cite{santoyo2021barrier, cosner2023robust}).
Such results can be summarized in the following theorem, which has been slightly modified and is taken from~\cite{cosner2024bounding}.
See \cite{cosner2024bounding} for further reference.
We use this result for comparisons in subsequent sections.
\begin{thm}\label{lem: Cosner}
Consider system~(\ref{eq: sys}).
  Let $h:\mathbb{R}^n\rightarrow\mathbb{R}$ be a continuous function and $\mathcal{C}$ be its corresponding safe set in \eqref{eq: safe set}.
  For some $B>0$, suppose that the following condition holds:
  \begin{equation}
  \label{eq:bounded-h}
    h(x)\leq B, \text{ for } x \in \mathbb{R}^n.
  \end{equation}
  Suppose that for all $0\le k\le K$ and $x\in\mathcal{C}$, and some $\alpha\in(0,1)$, there exists $u\in\mathbb{R}^m$ such that
  \begin{align}
    \label{eq:DTCBF}
    \mathbb{E}[h(F(x,u)+w_k) \mid \mathcal{F}_k]\geq
      \alpha h(x)
    \end{align}
    where $w_k$ is a Gaussian disturbance of system~(\ref{eq: sys}).
  Then, the \( K \)-step exit probability of the system with an initial state $x_0\in\mathcal{C}$ is bounded as:
  \begin{equation}
    P(x_0, K)\le
    1-\alpha^K\frac{h(x_0)}{B}.
    \label{eq: Kstep bound Cosner1}
  \end{equation}
  Similarly, suppose that for all $x \in \mathcal{C}$ and some $\beta \ge 0$, there exists $u \in \mathbb{R}^m$ such that
  \begin{align}\label{eq:c-martingale}
    \mathbb{E}[h(F(x,u)+w_k) \mid \mathcal{F}_k]\geq h(x) + \beta.
  \end{align}
  Then,
  \begin{equation}
    P(x_0, K)\le
    1- \frac{h(x_0)-\beta K}{B}.
        \label{eq: Kstep bound Cosner2}
  \end{equation}    
\end{thm}
This theorem provides the upper bounds~(\ref{eq: Kstep bound Cosner1}) and~(\ref{eq: Kstep bound Cosner2}) of \(K\)-step exit probabilities in affine forms of $h(x_0)$.
Tighter bounds on the \(K\)-step exit probabilities can be obtained if the affine forms of $h(x_0)$ in~(\ref{eq: Kstep bound Cosner1}) and~(\ref{eq: Kstep bound Cosner2}) are replaced with appropriately designed nonlinear forms of $h(x_0)$. 
This study extends the approach of Theorem~\ref{lem: Cosner} in this way to derive enhanced bounds for the probabilities and relax the boundedness condition~(\ref{eq:bounded-h}) by introducing auxiliary functions.

\section{MAIN RESULT}\label{sec: main}

In this section, to improve the bounds on exit probabilities, we first derive conditions for synthesizing a safety controller for the system \eqref{eq:closed-loop} and the safe set \eqref{eq: safe set}.
The conditions provide upper bounds for the $K$-step exit probability in \eqref{eq: Kstep exit}.
We then adapt the derived condition to the synthesis of safety controller, in a manner referred to as \textit{safety filter} in~\cite{Ames:19}, where a safety controller is designed by modifying a given nominal controller.
Subsequently, we demonstrate several specific choices of function $h$ such that safe controller synthesis can be analytically formulated as optimization problems.

As in standard safety-critical control problems, this study employs the function $h$ in~(\ref{eq: safe set}) as a CBF.
To develop safety controller synthesis with flexible bounds of \(K\)-step exit probability, we adopt the following definition of CBFs.
\begin{defn}
\label{defn:CBF}
Consider system~(\ref{eq: sys}).
Let \( h: \mathbb{R}^n \to \mathbb{R} \) be a continuous function, and the safe set $\mathcal{C}$ be determined by $h$ in the form of~(\ref{eq: safe set}).
Let $\Phi: \mathbb{R} \times \mathbb{Z}_{\ge 0} \to \mathbb{R}$ be a continuous function such that $\Phi(h,k)$ is decreasing in $h$ for all $0 \le k \le K$ and that $\Phi(h, k) \ge 0$ for $h \in \mathbb{R}$ and for all $0 \le k \le K$.
The function $h$ is the \textit{control barrier function (CBF) for \(K\)-step safety of system~(\ref{eq: sys}) with the auxiliary function $\Phi$} if, for all $0\leq k< K$ and $x \in \mathcal{C}$, there exists $u \in \mathbb{R}^m$ such that
\begin{equation}
    \mathbb{E} \left[ \Phi(h(F(x,u)+ w_k, k+1)) \mid \mathcal{F}_k \right] \le \Phi(h(x),k)\label{eq: StoDTCBF}
\end{equation}
holds where  $w_k$ is Gaussian disturbance of system~(\ref{eq: sys}). 
\end{defn}
In Definition~\ref{defn:CBF}, we introduce the auxiliary function $\Phi$.
As will be shown later, this function $\Phi$ enables to derive improved bounds for \(K\)-step exit probability in the synthesis of safety controllers.
In the following, when referring to a CBF in the sense of Definition~\ref{defn:CBF}, we may simply refer to it as a CBF rather than a CBF with the auxiliary function $\Phi$, when no confusion arises.

The following theorem is a key result providing the CBF characterization for the upper bound of \(K\)-step exit probability.
\begin{thm}\label{thm ville}
Consider system~(\ref{eq: sys}).
Suppose that a CBF for \(K\)-step safety of the system~(\ref{eq: sys}) with an auxiliary function $\Phi$ exists, which is denoted by $h:\mathbb{R}^n \to \mathbb{R}$.
% Let $\Phi$ be a function ensuring that $h$ becomes the CBF for \(K\)-step safety of system~(\ref{eq: sys}), as in Definition~\ref{defn:CBF}.
  Then, the \( K \)-step exit probability of the system with an initial state $x_0\in\mathcal{C}$ can be bounded as:
  \begin{align}
    P(x_0, K) \leq \frac{\Phi(h(x_0), 0)}{\min_{0\leq k\leq K}\Phi(0, k)}.
    \label{eq: risk probability}
  \end{align}
\end{thm}
\begin{pf}
Notice that, since $\Phi(h, k)$ is decreasing in $h$, 
$h < 0$ implies that $\Phi(h, k) > \Phi(0, k)$ for $1\le k \le K$.
Let $\mathcal{F}_k$-stopping time $\tau (w) := \min \left\{ k \in \mathbb{Z}_{\ge 0}; x_k \notin \mathcal{C} \right\}$, with $\min \emptyset = \infty$.
% We consider the stopped process $\Phi(h(x_{k\wedge \tau}), k\wedge\tau)$.
Then,
\begin{equation}
\label{eq:cond-1}
\mathbb{P} \left\{ \min_{0\leq k\leq K} h(x_k)<0 \right\} = \mathbb{P} \left\{ \mathcal{A}_1 \right\},
\end{equation}
where 
\begin{align}
  \mathcal{A}_1 &= \{ \omega \in \Omega \mid  \Phi(h(x_{k\wedge \tau}), k\wedge\tau)>\Phi(0, k\wedge \tau),\notag\\
  &\hspace{11em} \text{ for $0 \le k \le K$} \}.
\end{align}
Denoting the event
\begin{equation}
  \mathcal{A}_2 =\left\{ \omega \in \Omega \mid\! \max_{0\leq k\leq K}\!\Phi(h(x_{k\wedge \tau}), k\wedge\tau)>\!\min_{0\leq k\leq K}\!\Phi(0, k)  \right\},
\end{equation}
we obtain that $\mathcal{A}_1 \subset \mathcal{A}_2$, as the condition of $\mathcal{A}_1$ implies that of $\mathcal{A}_2$.
This implies 
\begin{equation}
\label{eq:cond-2}
\mathbb{P} \left\{ \mathcal{A}_1 \right\}
\le \mathbb{P} \{\mathcal{A}_2\}.
\end{equation}
Condition~(\ref{eq: StoDTCBF}) implies that there exists $u$ such that
\begin{equation}
\label{eq:super-martingale-condition}
\begin{aligned}
&\mathbb{E} \left[ \Phi(h(F(x_k, u)+ w_k, k+1)) \mid \mathcal{F}_k \right] \\
&{} = \mathbb{E} \left[ \Phi(h(x_{k+1}), k+1) \mid \mathcal{F}_k \right] \le \Phi(h(x_k),k)
\end{aligned}
\end{equation}
holds if $x_{k} \in \mathcal{C}$ at time $k$ and such control $u$ is applied to system~(\ref{eq: sys}), where we use equation~(\ref{eq: sys}) to derive the first equality.
This implies that $\Phi(h(x_{k \wedge \tau}),k \wedge \tau)$ is a nonnegative supermartingale over $0 \le k \le K$.
Furthermore, (\ref{eq:super-martingale-condition}) also implies that $\Phi(h(x_{k \wedge \tau \wedge K}),k \wedge \tau \wedge K)$ is a nonnegative supermartingale for $k \in \mathbb{Z}_{\ge 0}$.
Note that 
$\max_{0\leq k\leq K}\Phi(h(x_{k\wedge \tau}), k\wedge\tau) = \sup_{k \in \mathbb{Z}_{\ge 0}}\!\Phi(h(x_{k\wedge \tau \wedge K}), k\wedge\tau \wedge K)$ holds.
Then, for the event
\begin{equation}
\begin{aligned}
&\mathcal{A}_3 =\\
&\left\{ \omega \in \Omega \! \mid \! \sup_{k \in \mathbb{Z}_{\ge 0}}\!\Phi(h(x_{k\wedge \tau \wedge K}), k \! \wedge \! \tau \! \wedge \! K)> \min_{0 \le k \le K}\!\Phi(0, k)  \right\},
\end{aligned}  
\end{equation}
it holds that
\begin{equation}
\label{eq:condB-condC}
\mathbb{P} \left\{ \mathcal{A}_2 \right\} = \mathbb{P} \left\{ \mathcal{A}_3 \right\}.
\end{equation}
By applying Ville's inequality with $W_{k} = \Phi(h(x_{k \wedge \tau \wedge K}),k \wedge \tau \wedge K)$ and $\lambda = \min_{0 \le k \le K}\Phi(0, k)$ in~(\ref{eq:ville-inequality}), we obtain
\begin{equation}
\mathbb{P} \left\{ \mathcal{A}_3 \right\} \le \frac{\Phi (h(x_0), 0)}{\min_{0\leq k\leq K}\Phi(0, k)} \label{eq: theorem proof}.
\end{equation}
Combining~(\ref{eq:cond-1}), (\ref{eq:cond-2}), (\ref{eq:condB-condC}), and~(\ref{eq: theorem proof}) yields~(\ref{eq: risk probability}), which completes the proof.
\end{pf}

In the following, we focus on developing the safety controller synthesis based on Theorem~\ref{thm ville}.
A typical situation is that given a nominal controller $\{u_k^{\mathrm{nom}}\}_{0 \le k \le K-1}$, which is not necessarily one ensuring the \(K\)-step safety, we modify the controller so that the closed-loop system ensures the safety.
The following result presents a condition for the safety filter synthesis, which immediately follows from Theorem~\ref{thm ville}.
\begin{cor}
\label{cor:controller-design-condition}
Consider system~(\ref{eq: sys}) and the safe set $\mathcal{C}$ given by~(\ref{eq: safe set}) with a CBF $h: \mathbb{R}^n \to \mathbb{R}$ with an auxiliary function $\Phi$.
Given a nominal controller $\left\{ u_k^{\mathrm{nom}} \right\}_{0 \le k \le K-1}$, consider the minimization problem:
\begin{equation}
\label{eq:safety-filter}
% \begin{aligned}
  u_k^{\ast} = \mathrm{arg} \min_u \|u - u_k^{\mathrm{nom}}\|^2 \text{  s.t.~(\ref{eq: StoDTCBF}).}
% \end{aligned}  
\end{equation}
If the solution $u_k^{\ast}$ to minimization problem~(\ref{eq:safety-filter}) exists given $x = x_k$ in~(\ref{eq: StoDTCBF}) at every $0 \le k \le K-1$, $u_{k}^{\ast}$ ensures the upper bound of the \(K\)-step exit probability given by~(\ref{eq: risk probability}).
\end{cor}
Note that we obtain a feedback controller $u_k^{\ast} = k_c(x_k)$ for closed-loop system~(\ref{eq:closed-loop}) as the solution to~(\ref{eq:safety-filter}).

% Theorem~\ref{lem: Cosner} provides bounds on the $K$-step exit probability under the expectation conditions in (8). However, the bounds on the exit probability are relatively weak, as they are, at best, linear with respect to the value of $h\left(x_0\right) / B$.
% Additionally, the requirement for an upper bound on $h(x)$ limits the possible structures of $h$. For instance, affine functions cannot be used as $h$ in Theorem~\ref{lem: Cosner}.
% This study improves the upper bounds of \(K\)-step exit probabilities by using Theorem~\ref{thm ville} under tightened conditins on $h$. 
% % This study improves the upper bounds of \(K\)-step exit probabilities by using Theorem~\ref{thm ville} relaxing conditions on $h$. 
% In what follows, we show the safety controller synthesis condition using upper bounded functions $h$ first, followed by that using functions $h$ not necessarily bounded.

In what follows, we show specific choices of functions $h$ and $\Phi$ to develop safety controller synthesis conditions based on Theorem~\ref{thm ville}.
We first show a synthesis using upper bounded functions $h$, followed by that using functions $h$ not necessarily bounded.

\subsection{Bounded $h$}\label{subsec: bounded}

This section shows that for upper bounded $h$, a tighter bound of the \(K\)-step exit probability can be obtained by choosing specific functions $\Phi$ and $h$ in Theorem~\ref{thm ville}, compared to those in Theorem~\ref{lem: Cosner}.
In Theorem~\ref{lem: Cosner}, $\Phi$ is chosen as linear functions $\Phi(h,k)=\alpha^{-K}B-\alpha^{-k}h$ and $\Phi(h,k)=B-h+(K-k)\beta$ for \eqref{eq: Kstep bound Cosner1} and \eqref{eq: Kstep bound Cosner2}, respectively.
To tighten the \(K\)-step exit probability, as can be seen from (\ref{eq: risk probability}), $\Phi$ should be chosen so that $\Phi(h(x_0), 0)$ remains small even for small values of $h$.
Additionally, the computation of $\mathbb{E}[\Phi(h(F(x, u)+w_{k}),k+1) \mid \mathcal{F}_k]$ in~(\ref{eq: StoDTCBF}) must be analytically tractable as it is used in optimization problem~(\ref{eq:safety-filter}) to obtain safety controllers.
Apart from linear functions, few choices of $h$ and $\Phi$ satisfy both requirements.
Considering polynomial $h$ and $\Phi$ is one possible approach:
\begin{prop}\label{lem poly}
  Consider system~(\ref{eq: sys}).
  Let \( h: \mathbb{R}^n \to \mathbb{R} \) be a polynomial function, with $h(x)\leq B$ for all $x\in\mathbb{R}^n$.
  Let $\Psi(s):\mathbb{R}\rightarrow\mathbb{R}$ be a polynomial function, which is non-negative, continuous, and decreasing for $s\le0$.
  Suppose that for all $0\leq k< K$ and $x \in \mathcal{C}$, there exists $u \in \mathbb{R}^m$ such that 
  \begin{align}
    \mathbb{E}[\Psi(h(F(x,u)+w_k)-B) \mid \mathcal{F}_k]\leq \Psi(h(x)-B)+\beta.\label{eq: StoDTCBF polynomial}
  \end{align}
  where $w_k$ is the Gaussian disturbance of system~(\ref{eq: sys}).
  Then, 
  $h$ is a CBF for \(K\)-step safety of system~(\ref{eq: sys}) with the auxiliary function $\Phi$ given by
  \begin{equation}
  \label{eq:Phi-bounded-case}
  \Phi(h,k) = \Psi(h - B) + (K-k) \beta, \ h \in \mathbb{R}, \ 0 \le k \le K,
  \end{equation}
  and 
  the \( K \)-step exit probability of the system with an initial state $x_0\in\mathcal{C}$ is bounded as:
  \begin{align}
    \label{eq: K step polynomial}
    P(x_0, K) \leq \frac{\Psi(h(x_0)-B)+K\beta}{\Psi(-B)}.
  \end{align}
\end{prop}
\begin{pf}
  This proposition directly follows from Theorem \ref{thm ville}, by taking $\Phi$ as in~(\ref{eq:Phi-bounded-case}).
\end{pf}

Note that the upper bound of the \(K\)-step exit probability in~(\ref{eq: K step polynomial}) is given through function $\Psi$.
An appropriate choice of  $\Psi$ allows us to obtain tighter bound, compared with those of Theorem~\ref{lem: Cosner}.
If function $\Psi$ is polynomial function and $w_k$ is a Gaussian random variable, $\Psi(h(F(x,u) + w_k) - B)$ is integrable.
That is, $\mathbb{E} \left[ \Psi(h(F(x,u)+w_k) -B) \mid \mathcal{F}_k \right]$ has a finite value.
Furthermore, the expectation can typically be derived in a closed form.

The following corollary presents an application of Proposition~\ref{lem poly} to the safety filter, where Corollary~\ref{cor:controller-design-condition} becomes convex optimization problem.
\begin{cor}
\label{cor:poly convex}
  Under the same conditions as in Proposition~\ref{lem poly}, 
  suppose that $h$ is a concave function, and $\Psi(s)$ is convex and decreasing for $s\le 0$.
  Then, the minimization problem~(\ref{eq:safety-filter}) under condition \eqref{eq: StoDTCBF polynomial} is a convex programming problem with respect to $u$.
\end{cor}
\begin{pf}
Since the term $\|u-u_k^{\text{nom}}\|^2$  in~(\ref{eq:safety-filter}) is obviously convex with respect to $u$, we only show that the condition~(\ref{eq: StoDTCBF polynomial}), corresponding to the original constraint~(\ref{eq: StoDTCBF}) in the minimization problem~(\ref{eq:safety-filter}), is convex with respect to $u$.
To this end, we first show that, under the conditions of the corollary, $\Psi(h(x) - B)$ is convex with respect to $x$.
Indeed, for any $x_1,\ x_2\in\mathbb{R}^n$ and $\lambda\in[0,1]$,
  \begin{align}
    &\lambda\Psi(h(x_1)-B)+(1-\lambda)\Psi(h(x_2)-B)\notag\\
    &\ \ge\Psi(\lambda h(x_1)+(1-\lambda)h(x_2)-B)\notag\\
    &\ \ge\Psi(h(\lambda x_1+(1-\lambda)x_2)-B),
  \end{align}
where the first inequality follows from the convexity of $\Psi(s)$, and the second inequality follows from the concavity of $h$ and the decreasing property of $\Psi(s)$.
  Since expectation preserve convexity, \eqref{eq: StoDTCBF polynomial} is a convex constraint with respect to $F(x,u)+w$.
  Furthermore, since $F(x,u)$ is affine in $u$, \eqref{eq: StoDTCBF polynomial} is convex with respect to $u$.
\end{pf}

\begin{rem}
  Proposition~\ref{lem poly} provides a similar result compared to \cite{santoyo2021barrier}.
  The method in \cite{santoyo2021barrier} bounds the region of $x$ to construct a nonnegative supermartingale, which is essentially equivalent to bounding $x$ to ensure an upper bound on $h$.  
  However, restricting the state space $x$ is not advisable, as calculating the expectation of a Gaussian distribution over a truncated domain typically does not yield a closed-form solution.  
  Notably, the safe controller synthesis method for an affine barrier function proposed in \cite[Sec 4.2]{santoyo2021barrier} does not result in a polynomial form.
\end{rem}

\subsection{Unbounded $h$}\label{subsec: unbounded}

When there are no upper bounds on $h$, $\Phi$ cannot be constructed in the manner presented in the previous section.  
Indeed, there are few nonnegative and continuous functions $\Phi(h, k)$ that are decreasing in $h \geq 0$ and have a closed form for $\mathbb{E}[\Phi(h(F(x, u)+w_{k}),k+1) \mid \mathcal{F}_k]$ for $0 \le k \le K$.
One choice for $\Phi(h, k)$ is an exponential function.  
When $h$ is a general quadratic function and the system is driven by additive Gaussian disturbances, the closed form of \eqref{eq: StoDTCBF} with exponential $\Phi$ can be obtained and a bound for \(K\)-step exit probabilities are obtained.

\begin{thm}\label{SDTCBF quadratic}
Let \( h: \mathbb{R}^n \to \mathbb{R} \) be a quadratic function
\begin{equation}
\label{eq:quad-func}
h(x)=x^\top A x+b^\top x + c
\end{equation}
with a symmetric matrix $A\in\mathbb{R}^{n\times n}$, $b\in\mathbb{R}^n$ and $c\in\mathbb{R}$.
  Suppose that for all $0\leq k< K$, $x \in \mathcal{C}$ and some $\beta \geq 0$, there exists $u \in \mathbb{R}^m$ such that 
  \begin{equation}
    h(F(x,u))-\Theta (x, u) \geq -\log( \exp(-h(x)) +\beta) - M \label{eq: quadratic SDTCBF}
  \end{equation}
  where 
  \begin{align}
    \Theta(x,u) = & \left(AF(x,u)+\frac{b}{2} \right)^\top \Lambda^{-1} \left(AF(x,u)+\frac{b}{2}\right),\\
    M = &\frac{1}{2}\log \det (I+2\Sigma A),
  \end{align}
  and $\Lambda = \left(\frac{1}{2}\Sigma^{-1}+A\right)$, with $A$ satisfying the positive-definiteness of $\Lambda$,
  and $\Sigma$ is a covariance matrix of Gaussian disturbance of system~(\ref{eq: sys}).
  Then,
  $h$ is the CBF for \(K\)-step safety with the auxiliary function
  \begin{equation}
  \label{eq:Phi-unbounded-case}
  \Phi(h,k) = \exp(-h) + (K -k) \beta, \ h \in \mathbb{R}, \ 0 \le k \le K
  \end{equation}
  and
  the \(K\)-step exit probability of the system with an initial state $x_0\in\mathcal{C}$ is bounded as:
  \begin{align}
    P(x_0, K) \leq \exp(-h(x_0))+K\beta .\label{eq: quad prob bound}
  \end{align}
\end{thm}
\begin{pf}
  We first show that \eqref{eq: quadratic SDTCBF} implies \eqref{eq: StoDTCBF} by choosing $h(x)$ and $\Phi(h,k)$ as in~(\ref{eq:quad-func}) and~(\ref{eq:Phi-unbounded-case}), respectively.
  Note that $\Phi(h,k)$ is a non-negative continuous function, and is decreasing in $h$ and $k$.
   The choices of the functions yield the expression of condition~\eqref{eq: StoDTCBF} by
   \begin{equation}
   \label{eq:CBF-condition}
  \mathbb{E} \left[ \exp \left( -h( F(x_k,u_k) +w_k) \right) \mid \mathcal{F}_k\right] \le \exp (-h(x_{k})) + \beta.
  \end{equation}
  given $x_k$ at time $k$ and this condition is used for determining $u_k$.
  The left-hand side can be rewritten as 
  \begin{equation}
  \label{eq:Gaussian-integral}
  \begin{aligned}
    &\quad \mathbb{E}[\exp(-h(F+w_k)) \mid \mathcal{F}_k] \\
    &{} = \exp(-h(F))\mathbb{E}[\exp(-w_k^\top Aw_k -(2 AF + b)^\top w_k) \mid \mathcal{F}_k]
  \end{aligned}
  \end{equation}
  where we omit the arguments of function $F := F(x_k,u_k)$ for notational simplicity.
  To obtain the above equation, we use the property of the conditional expectation to move $\exp(-h(F))$ outside the expectation term since $x_{k}$ and $u_k$ are $\mathcal{F}_k$ measurable.
  The expectation in~(\ref{eq:Gaussian-integral}) is computed by using the Gaussian integral as follows:
  \begin{align}
    \mathbb{E}&[\exp(-\omega_k^\top A\omega_k -(2 AF+ b)^\top \omega_k) \mid \mathcal{F}_k]\notag\\
    &=\int_{\mathbb{R}^n}\frac{\exp(\!-\!\frac{1}{2}\omega_k^\top \Sigma^{-1}\omega_k\!-\!\omega_k^\top A\omega_k \!-\!(2 AF \!+\! b)^\top \omega_k)}{\sqrt{(2\pi)^n \det \Sigma}}d\omega_k\notag\\
    &=\frac{\exp(\Theta(x_k,u_k))}{\sqrt{(2\pi)^n \det \Sigma}}\int_{\mathbb{R}^n}\exp\left(\!-\!\frac{(\omega_k\!+\!\eta_k)^\top 2\Lambda(\omega_k\!+\!\eta_k)}{2} \right)\!dw_k\notag\\
    &=\frac{\exp(\Theta(x_k,u_k))}{\sqrt{(2\pi)^n \det \Sigma}}\sqrt{\frac{(2\pi)^n}{\det 2\Lambda}}=\exp(\Theta(x_k,u_k))M^{-\frac{1}{2}},\notag
  \end{align}
  with
  \begin{align*}
    \eta_k = \Lambda^{-1} \left( AF(x_k,u_k)+\frac{b}{2} \right).
  \end{align*}
  In the above derivation, again, we use the fact that $x_k$ and $u_k$ are measurable with respect to $\mathcal{F}_k$ to extract $\exp(\Theta(x_k, u_k))$ term from the conditional expectation, as $u_k$ is assumed to be adapted to $\mathcal{F}_k$.
  Note that according to the condition of the theorem, $A$ ensures the positive-definiteness of $\Lambda$, and thus $\det (I+2\Sigma A)>0$, so that the integral is finite and $M$ exists.
  Accordingly, condition~(\ref{eq: StoDTCBF}) is expressed as
  \begin{align}
    % &\mathbb{E}[\exp(-h(F(x_k, u_k)+w))]\leq \exp(-h(x_k))+\beta\label{eq: original quadratic SDTCBF}\\
    % \Leftrightarrow\ &\exp(-h(F))\mathbb{E}[\exp(-w^\top Aw -(2 AF + b)^\top w)]\notag\\ 
    &\ \exp(-h(F)+\Theta(x_k, u_k))(\det (I+2\Sigma A))^{-\frac{1}{2}} \notag\\
    &\ =\exp(-h(F)+\Theta(x_k, u_k)-\Phi)\notag\\
    &\ \leq \exp(-h(x_k))+\beta.\notag
  \end{align}
  By taking the logarithm of both sides and replacing $x_k$ and $u_k$ with $x$ and $u$, respectively, we obtain \eqref{eq: quadratic SDTCBF}.
  This implies that the function $h$ is the CBF in the sense of Definition~\ref{defn:CBF} with the auxiliary function $\Phi$ given by~(\ref{eq:Phi-unbounded-case}).

  From $\Phi(h(x_0),0)=\exp(-h(x_0))+K\beta$ and $\Phi(0,K)=1$, we obtain the probability bound \eqref{eq: quad prob bound} based on Theorem~\ref{thm ville}.
  This completes the proof.
\end{pf}

Note that Theorem \ref{SDTCBF quadratic} includes the case where $h$ is affine.
By substituting $A=O_{n\times n}$ to Theorem \ref{SDTCBF quadratic}, constraint \eqref{eq: quadratic SDTCBF} for affine CBF, $h(x)=b^\top x +c$, can be shown as follows:
\begin{equation}
  \begin{aligned}
    h(F(x,u)) \geq -\log(\exp(-h(x))+\beta)+\frac{1}{2}b^\top \Sigma b \label{eq: affine SDTCBF}
  \end{aligned}
\end{equation}

We show the following corollary as a special case of Corollary~\ref{cor:controller-design-condition} for the safety filter, which yields a condition that the minimization problem~(\ref{eq:safety-filter}) becomes convex.
% The convexity facilitates the computation of the safety filter.
\begin{cor}
\label{cor-quadratic}
Under the same conditions as in Theorem~\ref{SDTCBF quadratic}, 
suppose that the matrix $A$ and the covariance matrix $\Sigma$ are such that
\begin{equation}
\label{eq:matrix-M}
% N = A-A \left( \frac{1}{2}\Sigma^{-1}+A \right)^{-1}A
N = A-A \Lambda^{-1}A
\end{equation}
is negative semi-definite.
Then, the minimization problem~(\ref{eq:safety-filter}) is a convex programming problem with respect to $u$.
Particularly, if $A$ is a negative semidefinite matrix, $N$ is a negative semidefinite matrix.
\end{cor}
\begin{pf}
Under the setting of Proposition~\ref{SDTCBF quadratic}, condition~(\ref{eq: quadratic SDTCBF}) implies constraint~(\ref{eq: StoDTCBF}) in the minimization problem~(\ref{eq:safety-filter}).
In condition~(\ref{eq: quadratic SDTCBF}), the matrix $N$ appears as the coefficient matrix of the quadratic term with respect to $g(x)u$.
Accordingly, with $N$ given by~(\ref{eq:matrix-M}), (\ref{eq: quadratic SDTCBF}) becomes convex with respect to $u$.
With the fact that the objective function $\|u - u_k^{\text{nom}}\|^2$ is convex with respect to $u$ and the fact that the constraint (\ref{eq: StoDTCBF}) in the minimization problem~(\ref{eq:safety-filter}) is ensured by~(\ref{eq: quadratic SDTCBF}), the problem~(\ref{eq:safety-filter}) becomes the convex programming problem with respect to $u$.
Lastly, if the matrix $A$ is a negative semidefinite matrix, the definition of $N$ in~(\ref{eq:matrix-M}) implies that $N$ becomes a negative semidefinite matrix with the fact that $\Lambda$ is a positive definite matrix under the conditions of the corollary.
\end{pf}
% By choosing $A$ as a negative semi-definite matrix, $N$ always becomes negative semi-definite.
Note that Corollary~\ref{cor-quadratic} holds for the bounded case and the affine CBF case.

Lastly, we introduce a technique to obtain a tighter bound on \(K\)-step exit probability by modifying a CBF $h$.
This technique replaces a given CBF $h(x)$ with the scaled CBF $a h(x)$ with the scaling parameter $a \ge 1$.
Using $ah(x)$ instead of $h(x)$ in condition~(\ref{eq: quadratic SDTCBF}) makes the bound arbitrarily tighter.
Note that the scaling operation does not change the safe set defined by $h(x)$.
At the same time, the introduction of $a$ tightens the constraint~(\ref{eq: quadratic SDTCBF}) for determining the control $u$.
This can be seen from~(\ref{eq:CBF-condition}); the term $\exp(- a h(F(x,u) + w_k))$, instead of $\exp(- h(F(x,u) + w_k))$, takes a larger value when $h(F(x,u)+w_k)$ is negative.
% This can be seen from~(\ref{eq:CBF-condition}); the term $\exp(- a h(F(x,u) + w))$, instead of $\exp(- h(F(x,u) + w))$, takes a larger value for a smaller value of $w$, compared with the condition using $\exp(- h(F(x,u) + w))$.
This affects the expectation $\mathbb{E} \left[ \exp \left( -h( F(x,u) +w_k) \mid \mathcal{F}_k \right) \right]$ in~(\ref{eq: StoDTCBF}) and tightens the constraint with respect to $u$.

\begin{rem}
  Theorem \ref{SDTCBF quadratic} gives a similar result to \cite{steinhardt2012finite}, when $h(x)=x^\top A x + c$ with negative definite $A$ and $c>0$ are used.
  It is essentially the same when system dynamics are polynomials and when we choose $\Phi = \exp(-h)+(K-k)\beta-\exp(-c)$.
  Theorem \ref{SDTCBF quadratic} is a generalized condition, including unbounded $h$.
\end{rem}

\section{NUMERICAL EXAMPLES}\label{sec: num}
This section presents numerical examples to demonstrate the effectiveness of the safety filter in Corollary~\ref{cor:controller-design-condition}, when CBFs for $K$-step safety of system~\eqref{eq: sys} with several different auxiliary function $\Psi$ are used.

\subsection{affine CBF}
\label{sec: affine CBF}
We begin with a simple example of safety controller synthesis using Theorem~\ref{SDTCBF quadratic}.
Consider the following discretized model of the continuous-time linear one-dimensional system $\dot{x} = u$ perturbed by Gaussian noises, with an affine CBF $h(x)=x$ and its corresponding safety set $\mathcal{C}$:
\begin{align}
  x_{k+1} &= x_k + u_k \, \Delta t + \omega_k, \quad \omega_k \sim \mathcal{N}(0, \sigma^2 \Delta t) \\
  \mathcal{C} &= \{x \mid h(x) \geq 0 \}. \label{eq: safe set affine}
\end{align}
where $\Delta t$ is the time step size in the discretization.
We consider the safe controller synthesis with the nominal control input $u_k^{\rm nom} = -x_k$.

\begin{figure}[t]
  \centering
  \includegraphics[width = 8.8cm]{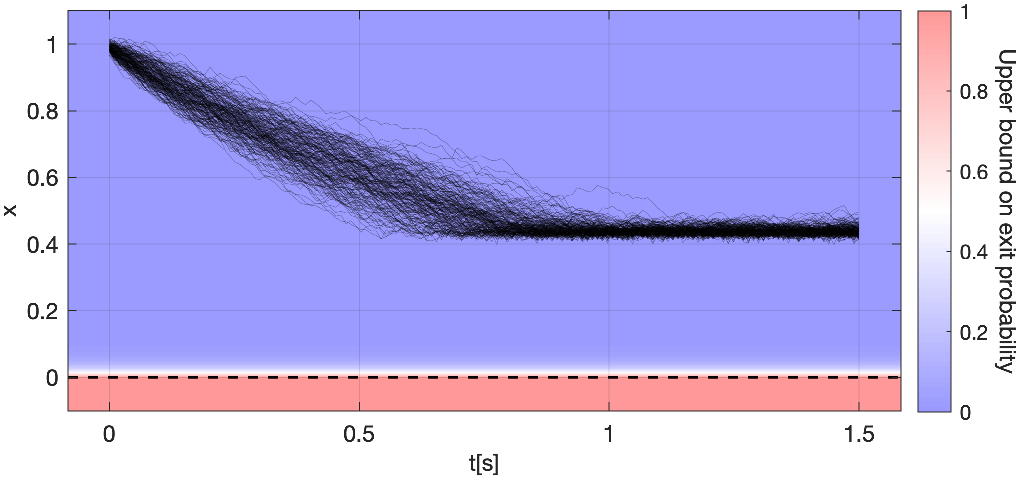}
  \caption{Safe control with an affine CBF over 200 trials. The color map shows the bounds of $P(x, 150)$ for each state $x$.\label{fig: affine}}
\end{figure}

Using the aforementioned technique replacing the function $h(x)$ with the scaled function $a h(x)$, the condition~(\ref{eq: quadratic SDTCBF}) in Theorem~\ref{SDTCBF quadratic} is expressed as 
\begin{equation}
  \begin{aligned}
    a(x+u\Delta t) \geq \log(\exp(-ax)+\beta)+\frac{a^2}{2}\sigma^2\Delta t. 
  \end{aligned}
\end{equation}
For this simulation, we set the parameters as follows: 
\begin{align*}
    a = 50,\
    \beta = 10^{-4},\
    \Delta t = 0.01, \
    \sigma = 1, \
    x_0 = 1.
\end{align*}
Fig.~\ref{fig: affine} shows the result of the simulation.
This figure shows 200 sample paths with the safety controller.
It can be observed that the paths remains within the safe region, where the exit probability over 150 steps remains low.
The heat map in the background of Fig.~\ref{fig: affine} shows the values of the upper bound of the \(K\)-step exit probability $P(x, K)$ for each $x$ with $K=150$.
Notably, the upper bound of the exit probability is small even if $x$ is close to the boundary of the safe set, $x=0$, which indicates that Theorem~\ref{SDTCBF quadratic} provides a tight bound of the exit probability.

\subsection{Inverted Pendulum}
\label{sec:inverted-pendulum}
We next consider the following case with bounded CBF, a discretized model of inverted pendulum about its upright position:
\begin{equation}
  x_{k+1}=
  \begin{bmatrix}
    \theta_{k+1}\\\dot{\theta}_{k+1}
  \end{bmatrix}
  =\underbrace{
  \begin{bmatrix}
    \theta_k+\Delta t \dot{\theta}_k\\\dot{\theta}_k+\Delta t \sin(\theta_k)
  \end{bmatrix}
  +
  \begin{bmatrix}
    0\\ \Delta t u_k
  \end{bmatrix}}_{F(x_k,u_k)}
  +w_k,
\end{equation}
with $\Delta t = 0.01$, $\omega_k \sim \mathcal{N}(0, \Sigma \Delta t)$ with $\Sigma = {\rm diag}\left(0.05^2, 0.25^2\right)$, and a quadratic CBF $h(x)=x^\top A x+1$ with
\begin{equation}
  A = -\frac{6^2}{\pi^2}
  \begin{bmatrix}
    1 & 3^{-\frac{1}{2}}\\3^{-\frac{1}{2}}&1
  \end{bmatrix}.
\end{equation} 
Here, the system dynamics and the safety set are adapted from \cite{cosner2023robust}.

We compare the safety bounds obtained by condition~\eqref{eq:DTCBF} in Theorem~\ref{lem: Cosner}, condition~\eqref{eq: StoDTCBF polynomial} in Proposition~\ref{lem poly}, and condition \eqref{eq: quadratic SDTCBF} in Theorem~\ref{SDTCBF quadratic}.
Condition \eqref{eq:DTCBF} and the $K$-step exit probability~(\ref{eq: Kstep bound Cosner1}) can then be rewritten as 
\begin{align}
  & h(F(x,u))+{\rm Tr}(A\Sigma)\ge \alpha h(x), \label{eq: cosner pendulum}\\
  & P(x_0, K)\le 1-\alpha^Kh(x_0),
\end{align}
respectively, where we set $\alpha = 1+{\rm Tr}(A\Sigma)$, the highest value it can take to guarantee the feasibility of \eqref{eq: cosner pendulum} when $h(x)=1$.\\
For Proposition~\ref{lem poly}, we use $\Psi(x)=x^2$.
Then, after performing some calculations to derive the explicit form of~(\ref{eq: StoDTCBF polynomial}), the condition~\eqref{eq: StoDTCBF polynomial} and the $K$-step exit probability~\eqref{eq: K step polynomial} can be rewritten as
\begin{align}
  &(h(F)-1)^2+ 4F^\top A\Sigma AF+2F^\top A F{\rm Tr}(A\Sigma)\notag\\
  &\hspace{1em}+2 {\rm Tr}((A \Sigma)^2) + ({\rm Tr}(A \Sigma))^2\ge (h(x)-1)^2 + \beta \label{eq: polynomial pendulum}\\
  & P(x_0, K)\le (h(x_0)-1)^2+\beta K,
\end{align}
respectively, where $F=F(x,u)$.
For \eqref{eq: polynomial pendulum}, we set $\beta = {\rm Tr}((A \Sigma)^2) + ({\rm Tr}(A \Sigma))^2$, which is the maximum value which ensures the feasibility of \eqref{eq: polynomial pendulum} for all $x\in\mathcal{C}$.
This choice is also justified by examining the case where $x=0$ and $F(x,u)=0$.
Note that \eqref{eq: polynomial pendulum} gives a convex constraint with respect to $u$, as also shown in Corollary~\ref{cor:poly convex}.\\
For Theorem~\ref{thm ville}, similarly to Subsection~\ref{sec: affine CBF}, we introduce a scalar parameter $a$ to tighten the exit probability.
Then, \eqref{eq: quadratic SDTCBF} and the $K$-step exit probability \eqref{eq: quad prob bound} can be rewritten as 
\begin{align}
  &ah(F)- F^\top aA \left( \frac{1}{2}\Sigma^{-1}+aA \right)aAF \notag\\
  &\ \ \ge -\log(\exp (-ah(x))+\beta)-\frac{1}{2}\log\det(I+2\Sigma aA) \label{eq: quad pendulum}\\
  & P(x_0, K)\le \exp(-ah(x_0))+ K\beta,
\end{align}
respectively, where $F=F(x,u)$.
We set $a = 10$ and $\beta=10^{-5}$, values that ensure the feasibility of \eqref{eq: quad pendulum} for the case $x=0$ and $F(x,u)=0$, and consequently for all $x\in\mathcal{C}$.
Note that \eqref{eq: quad pendulum} also gives a convex constraint with respect to $u$, as shown in Corollary~\ref{cor-quadratic}.\\

\begin{figure}[t] % figのlemmaの番号は提出前に変えます
  \centering
  \includegraphics[width = 8.8cm]{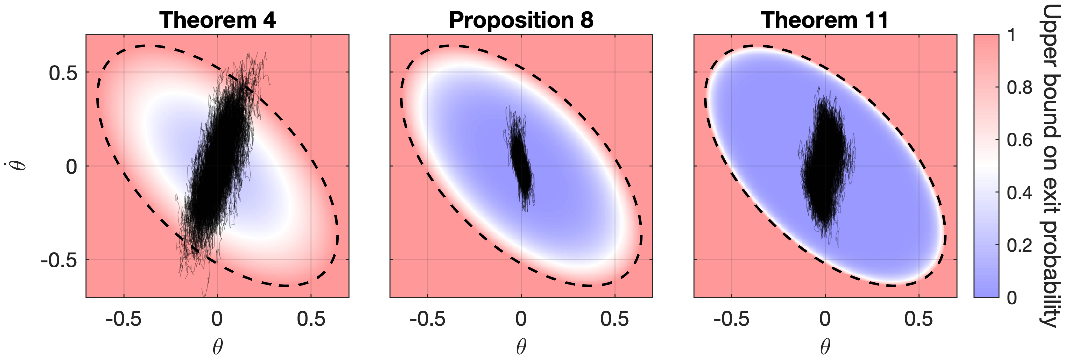}
  \caption{Safe control of an inverted pendulum using different conditions for 500 trials. The dotted lines show the boundary of the safe set $h(x)=0$, and the color map shows the bounds of $P(x, 100)$ for each state $x$. \label{fig: pendulum}}
\end{figure}

The simulation results with $x_0=[0,0]^\top$, $u_k^{\rm nom}=0$ and $K=100$ for 500 trials of each method are shown in Fig.~\ref{fig: pendulum}.
Upperbounds of $P(x_0, 100)$ for Theorem~\ref{lem: Cosner}, Proposition~\ref{lem poly} and Theorem~\ref{SDTCBF quadratic} are approximately $21.1\%$, $0.16\%$, and $0.10\%$, respectively.
It can be observed that Theorem~\ref{SDTCBF quadratic} yields the tightest bound, followed by Proposition~\ref{lem poly}.

\subsection{Single Integrator Obstacle Avoidance}
Lastly, we consider the control of a discretized model of 2D unit-mass single-integrator dynamics, avoiding obstacles.
This is a discretized model of the case considered in \cite{singletary2021comparative}.
The system dynamics is given as 
\begin{equation}
  x_{k+1}=
  x_k + u_k\Delta t + w_k,
\end{equation}
with $\Delta t = 0.01$, $w_k \sim \mathcal{N}\left(0_2, {\rm diag}\left(0.02\Delta t, 0.02 \Delta t\right)\right)$.
We consider the case where a quadratic CBF $h(x)=x^\top A x + b^T x + c$ yields an unbounded safe set $\mathcal{C}$.
For the safe controller synthesis \eqref{eq:safety-filter} with conditions \eqref{eq: quadratic SDTCBF} in Theorem~\ref{SDTCBF quadratic}, similarly to Subsections~\ref{sec: affine CBF} and \ref{sec:inverted-pendulum}, we introduce a scalar parameter $a$ to tighten the exit probability.

For the nominal controller, we use a simple proportional controller with a gain of $1$ on position, defined as  
\begin{equation}
  u_k^{\rm nom} = -(x_k - x_{\rm goal}),
\end{equation}  
where \(x_{\rm goal} \in \mathbb{R}^2\) represents the desired destination.

In this example, we consider safety-critical control using a single CBF as well as multiple CBFs to address complex safe sets.

\subsubsection{Single CBF}
\begin{figure}[t]
  \centering
  \includegraphics[width = 8.8cm]{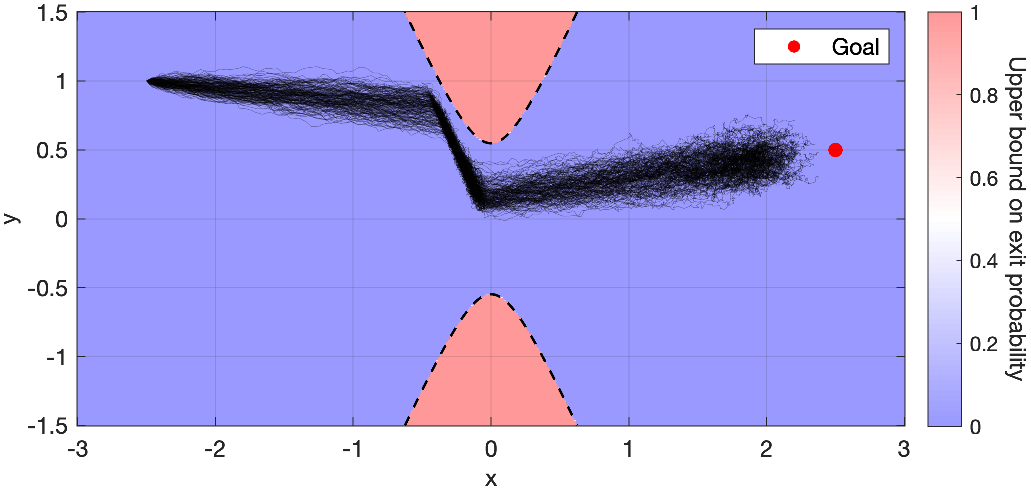}
  \caption{Safe control of a single integrator model using Theorem~\ref{thm ville} for 200 trials. The dotted line shows the boundary of the safe set $h(x)=0$, and the color map shows the bounds of $P(x, 300)$ for each state $x$. \label{fig: hyparabora}}
\end{figure}

We first consider the case when $h(x)=0$ is a hyperbora, defined by $A={\rm diag}\left(5,-1\right)$, $b=0_2$, and $c=0.3$ in the quadratic CBF $h(x)$.
This case considers an unbounded safe set with unbounded barrier functions under Gaussian distributed noise, a scenario not addressed by the aforementioned existing methods.
Note that the condition \eqref{eq: quadratic SDTCBF} is not convex with respect to $u$, making \eqref{eq:safety-filter} a non-convex optimization problem.
The simulation results with $x_0=[-2.5, 1]^\top$, $x_{\rm goal}=[2.5, 0.5]^\top$, $a=20$, $\beta=10^{-4}$, and $K=300$ for 200 trials are shown in Fig.~\ref{fig: hyparabora}.
The upperbound for $P(x_0, 300)$ under this condition is approximately $3.00\%$.
It can be observed that the system remains inside the safe region in all cases, showing the efficacy of our proposed method.

\subsubsection{Multiple CBFs}
Next, we consider a scenario in which the system must avoid $l$ obstacles, each represented by a distinct CBF, $h_i(x)$, $i=1,\dots l$.
Then, the safe set $\mathcal{C}$ can be defined as
\begin{equation}
  \mathcal{C}=\{x\in\mathbb{R}^n \mid \ \forall i=1,\dots,l,\ h_i(x)\ge 0\}.
\end{equation}
In this case, we can evaluate the $K$-step exit probability bound for $\mathcal{C}$ by evaluating $K$-step exit probability bound for each subset of the safe set given by
\begin{equation}
  \mathcal{C}_i=\{x\in\mathbb{R}^n \mid \ h_i(x)\ge 0\}.
\end{equation}
Theorem~\ref{thm ville} yields $K$-step exit probability bound for each $\mathcal{C}_i$ by designing a function $\Phi_i$ as the auxiliary function of $h_i$, as
\begin{equation}
  \begin{aligned}
    P_{\mathcal{C}_i}\left(x_0, K\right) &:= \mathbb{P}\left(\min_{0\leq k\leq K}h_i(x_k)<0\right) \\
    &\le \frac{\Phi_i (h_i(x_0), 0)}{\min_{0\leq k\leq K}\Phi_i(0, k)}.
  \end{aligned}
\end{equation}  
  Then, by applying Boole's inequality, we obtain a bound of the \(K\)-step exit probability for the safe set $\mathcal{C}$ from the bounds of probabilities for $\mathcal{C}_i$, as 
\begin{align}
  & \mathbb{P} \left(\min_{0\leq k\leq K, i = 1,\dots l} h_i(x_k)<0 \right)=\mathbb{P}\left(\bigcup_{i=1}^l \min_{0\leq k\leq K}h_i(x_k)<0\right)\notag\\
  & \le\sum_{i=1}^l P_{\mathcal{C}_i} (x_0, K)
  % & \le\sum_{i=1}^N\mathbb{P}\left(\min_{0\leq k\leq K}h_i(x_k)<0\right)\notag\\
  \le\sum_{i=1}^l \frac{\Phi_i (h_i(x_0), 0)}{\min_{0\leq k\leq K}\Phi_i(0, k)}.\label{eq: multi prob bound}
\end{align}

We consider avoiding four circular obstacles, each with a radius of \(0.4\), centered at \([-1.5, 0.7]^\top\), \([0.5, 0.7]^\top\), \([-0.5, -0.7]^\top\), and \([1.5, -0.7]^\top\), respectively.
For the corresponding CBFs, we set $A_i=I_2$ and define $b_i, c_i$ for $i=1,\dots, 4$, accordingly.
Here, similary to the single CBF cases, it can be seen that the bound on the $K$-step exit probability can be tightened by introducing scalar parameters $a_i$ and consider $a_ih_i(x)$ in \eqref{eq: multi prob bound}.
We set $a_i=20$ and $\beta_i=10^{-5}$ for all $i=1,\dots 4$.

\begin{figure}[t]
  \centering
  \includegraphics[width = 8.8cm]{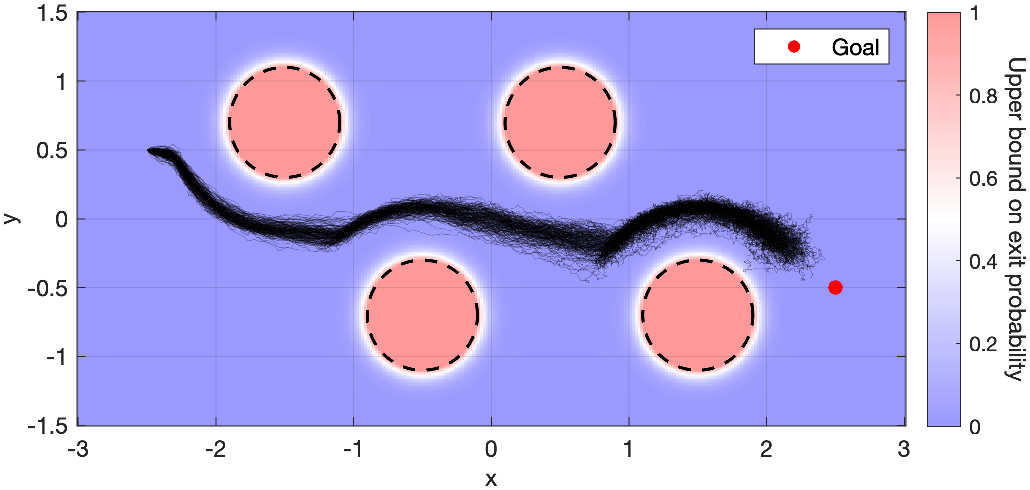}
  \caption{Safe control of a single integrator model using Theorem~\ref{thm ville} for 200 trials. The dotted line shows the boundary of the safe set $h(x)=0$, and the color map shows the bounds of $P(x, 300)$ for each state $x$. \label{fig: multiple cbfs}}
\end{figure}

The simulation results with $x_0=[-2.5, 0.5]^\top$, $x_{\rm goal}=[2.5, -0.5]^\top$, and $K=300$ for 200 trials are shown in Fig.~\ref{fig: multiple cbfs}.
The upperbound for $P(x_0, 300)$ under this condition is approximately $0.30\%$.
It can be observed that the system remains inside the safe region for all trials, showing the efficacy of our proposed method.

\section{CONCLUSION}
In this work, we presented conditions for probabilistically safe controller synthesis in stochastic systems to provide flexible bounds of the safe probability.
Future directions include exploring more constructive methods for determining parameters in the design of controllers, and
extending this approach to more complex scenarios such as continuous-time stochastic systems as in~\cite{hoshino2023scalable,nishimura2024control},
% exploring the relation to other approaches on safe probability bounds such as~\cite{liu2024safety},
and comparing with non-martingale-based finite-time safety guarantee method such as \cite{black2023safety} and~\cite{liu2024safety}.

\bibliography{main}             % bib file to produce the bibliography
\end{document}